\documentclass[twocolumn,showpacs,preprintnumbers,amsmath,amssymb]{revtex4}
\usepackage{tabularx,graphicx}

\usepackage{color}
\usepackage{hyperref}
\hypersetup{
    colorlinks=true,
    linkcolor=blue,
    filecolor=blue,      
    urlcolor=blue,
}

\begin{document}

\newcommand{\beq}{\begin{equation}}
\newcommand{\eeq}{\end{equation}}
\newcommand{\beqn}{\begin{eqnarray}}
\newcommand{\eeqn}{\end{eqnarray}}
\newcommand{\bmath}{\begin{subequations}}
\newcommand{\emath}{\end{subequations}}
\newcommand{\bra}[1]{\langle #1|}
\newcommand{\ket}[1]{|#1\rangle}

\title{  Joule heating in the normal-superconductor phase transition in a magnetic field}
\author{J. E. Hirsch }
\address{Department of Physics, University of California, San Diego,
La Jolla, CA 92093-0319}

\begin{abstract} 
Joule heating is a non-equilibrium dissipative process that occurs in a normal metal when an electric current flows,  in an amount proportional to  the
metal's resistance. When it is induced by eddy currents resulting from a change in magnetic flux, it is also proportional to  the rate at which the magnetic flux
changes. Here we show that in the phase transformation between normal and superconducting states of a metal in a magnetic field, the total amount of Joule heating 
is determined by  the thermodynamic properties of the system and is independent of the resistivity of the normal metal. 
We also show that Joule heating only occurs in the normal region of the material. The conventional theory of superconductivity however predicts
that Joule heating occurs also in the superconducting region within a London penetration depth of the phase boundary.  This implies that there is
a problem with the  conventional theory of superconductivity.
 \end{abstract}
\maketitle

            \begin{figure} 
 \resizebox{7.5cm}{!}{\includegraphics[width=6cm]{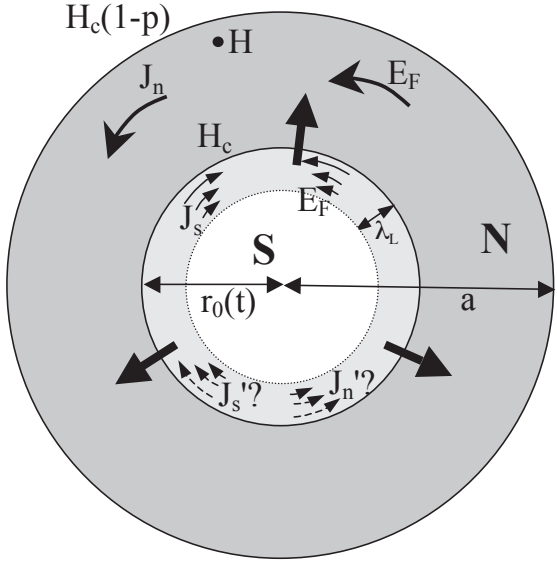}} 
 \caption { N-S transition. Cylindrical superconductor of unit height $h$ and radius $a<<h$ seen from the top. 
 Magnetic field $H$ points out of the paper. The phase boundary denoted by $r_0(t)$ is moving outward.
    In the normal region (dark grey), and in the superconducting region  within a London penetration depth $\lambda_L$   of the phase boundary
    (light grey)  a Faraday electric field $E_F$ pointing
 counterclockwise exists. $J_n$ and $J_s$ denote normal current and supercurrent respectively.  
 $J_n'$ and $J_s'$ are hypothesized normal current and additional supercurrent near the phase boundary.  }
 \label{figure1}
 \end{figure} 
\section{The problem}
Consider a cylindrical  type I superconductor in the presence of a uniform magnetic field along its axis undergoing
a transition from the normal (N) to the superconducting (S) state as shown in Fig. 1, or from the superconducting to the normal state as shown in Fig. 2. 
For simplicity we assume cylindrical symmetry throughout the process.
In the superconducting region within a London penetration depth $\lambda_L$  of the phase boundary, of radius  $r_0(t)$, a supercurrent $J_s$ flows that
nullifies the magnetic field in the interior. We will call that region   the `boundary layer' in what follows.
A  Faraday electric field $E_F$ exists throughout the
normal region $r\geq r_0(t)$  as well as in the boundary layer during the transition, that points counterclockwise in the N-S transition and clockwise in the S-N transition,
as shown in the figures. The Faraday electric field induces a normal current in the normal region during the transition process, that will
dissipate Joule heat.

             \begin{figure} 
 \resizebox{7.5cm}{!}{\includegraphics[width=6cm]{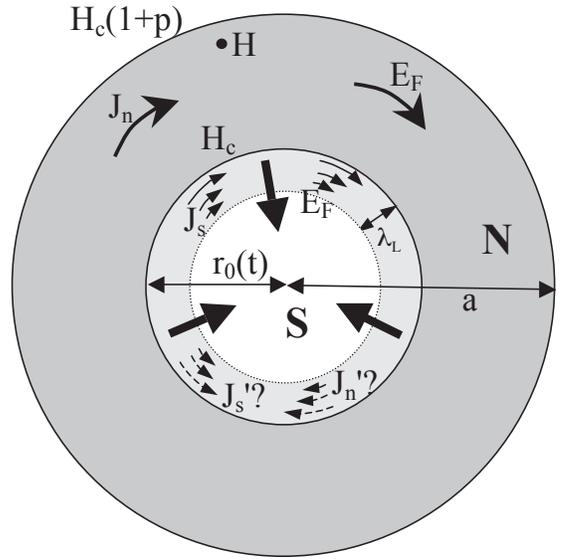}} 
 \caption { S-N transition. The phase boundary denoted by $r_0(t)$ is moving inward.   The Faraday electric field $E_F$ and normal current $J_n$  
 is  in opposite direction
 to Fig. 1, the supercurrent $J_s$ is in the same direction as in Fig. 1.   }
 \label{figure1}
 \end{figure} 
 
 In this paper we show that the total Joule heat dissipated is independent of the resistivity of the normal metal and of the rate at which the process
 occurs, and depends only on thermodynamic properties of the system. 
 To our knowledge, this has not been pointed out  in the literature before. In addition, we show  that thermodynamics requires that no Joule heat
 is dissipated in the boundary layer during the transition. We point out  that   the conventional theory of superconductivity \cite{tinkham} predicts the existence of such a normal current and associated Joule heat in the boundary layer.
Therefore we conclude that there is a problem with the  conventional theory of superconductivity. Instead, we point out that the problem does not
arise within the alternative theory of hole superconductivity \cite{holesc}.

In recent work we have shown that thermodynamic considerations for a superconductor in a magnetic field in a process where the temperature is changed between temperatures, both below $T_c$, lead to the same conclusion regarding the conventional theory of superconductivity \cite{inconsistency}. In other recent work we have shown that consideration of 
the entropy production associated with transfer of momentum   between electrons and the body during the transition
 between normal and superconductings states also  leads to the same conclusion \cite{entropy}.

            \begin{figure} 
 \resizebox{7.5cm}{!}{\includegraphics[width=6cm]{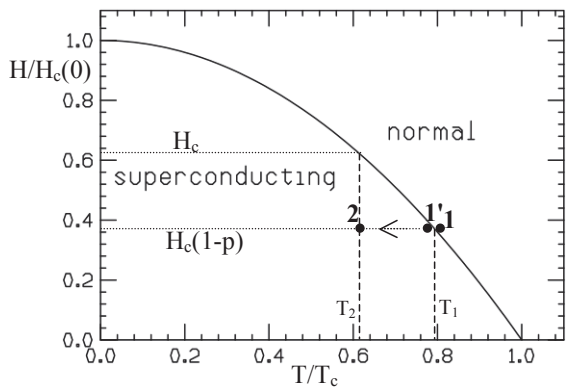}} 
 \caption { N-S transition: states 1 and  1' are normal and superconducting states on the coexistence curve at temperature $T_1$, state 2 is the superconducting state at lower
 temperature $T_2$, all in external field $H_c(T_1)\equiv H_c(1-p)$.  The critical field at temperature $T_2$ is $H_c\equiv H_c(T_2)$.
States 1, 1', and 2 are equilibrium states.
  The supercooled non-equilibrium state 2' is the normal state for the same temperature and magnetic field as the equilibrium superconducting state 2.
  The system interchanges heat with a heat reservoir at temperature $T_2$}
 \label{figure1}
 \end{figure} 
 
             \begin{figure} 
 \resizebox{7.5cm}{!}{\includegraphics[width=6cm]{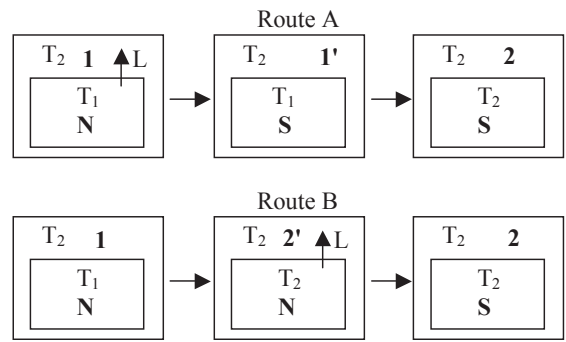}} 
 \caption {Two routes for the transition between equilibrium states 1 and 2 in figure 3 (see text). $L$ denotes latent heat.  
 The state 2' is a non-equilibrium state where  the system is in the supercooled normal state at temperature 
 $T_2$, magnetic field $H_c(1-p)$.}
 \label{figure1}
 \end{figure}

\section{Thermodynamics}
We consider first the N-S transition shown in Fig. 3, i.e. the Meissner effect, in an applied magnetic field $H_c(1-p)$. Here, 
$H_c\equiv H_c(T_2)$ and $H_c(1-p)\equiv H_c(T_1)$ are the critical magnetic fields at temperatures $T_2$ and $T_1>T_2$, and $p>0$. 
The system is initially in the normal state denoted by 1, at temperature infinitesimally above 
$T_1$.  In the final state 2, at temperature $T_2$, the system is in the superconducting state with the magnetic field excluded from its
interior. The state 1' shown in Fig. 3 denotes the system in the superconducting state at temperature infinitesimally below $T_1$
with the magnetic field excluded.
The three states 1, 1' and 2 are equilibrium states of the system.

Consider the two different routes between the same  initial and final equilibrium states 1 and 2 shown in Fig. 4, denoted by route A and route B. 

In route A, the system undergoes the normal-superconductor (N-S) transition at the coexistence curve at 
temperature $T_1$ transferring latent heat $L(T_1)$ to a reservoir at temperature $T_2$. 
Then, it cools to temperature $T_2$ transferring additional heat to the reservoir and coming into thermal equilibrium
with it. The transition proceeds infinitely slowly because it occurs on the coexistence curve, and no Joule heat is generated as the magnetic field is
expelled. The total heat  (per unit volume) transferred to the
reservoir is  
\beq
Q_A=L(T_1)+\int_{T_2}^{T_1} dT C_s(T)  .
\eeq
 Here $C_s \equiv C_s(T)$ is the heat capacity of the system in the superconducting state.
The change in entropy of the universe (system plus reservoir)  in this process   results
from the transfer of latent heat between the system and the
reservoir at different temperatures during the transition  as well as from the transfer of heat during the cooling of the system from $T_1$ to $T_2$:
\beq
\Delta S_{univ, A}=L(T_1)(\frac{1}{T_2}-\frac{1}{T_1})  + \int_{T_2}^{T_1} dT C_s(T)[\frac{1}{T_2}-\frac{1}{T}]  
\eeq

In route B, we assume the system in the normal state 1 at temperature $T_1$  is rapidly supercooled to  temperature
$T_2$ by contact with the heat reservoir while remaining in the normal state, then undergoes the transition to the superconducting state
while at temperature $T_2$, expelling the magnetic field in a finite amount of time hence   generating Joule heat in the process,  transferring both  the latent heat and the Joule heat to the reservoir at temperature $T_2$. 

The total heat transferred to the reservoir in route B is then
\beq
Q_B= \int_{T_2}^{T_1} dT C_n(T) +L(T_2)+Q_J
\eeq
where $C_n$ is the heat capacity of the system in the normal state  and $Q_J$ is the Joule heat generated in the transition from the supercooled normal state at temperature $T_2$  to the superconducting state
at temperature $T_2$, which takes a finite time.
The change of entropy of the universe in route B   is  due to the transfer of heat between the system and the reservoir during cooling of the system
in the normal state, and 
the generation of 
Joule heat during the transition. Since the latent heat and the Joule heat are transferred between system and reservoir at the same temperature $T_2$, this transfer does not change the
entropy of the universe. Hence the change in entropy of the universe in route B  is given by
\beq
\Delta S_{univ, B}=   \int_{T_2}^{T_1} dT C_n(T)[\frac{1}{T_2}-\frac{1}{T}]   +\frac{Q_J}{T_2}
\eeq

It is important to understand  that $both$ the final states of the system and of  the reservoir are the same in routes A and B,
whether the `reservoir' is infinite or finite.  For an infinite reservoir its temperature $T_2$ is
unchanged, as assumed here  for simplicity. For a finite `reservoir', it and the system will reach an equilibrium temperature 
$T_3$, with $T_2<T_3 < T_1$. 
If the  final equilibrium temperatures in the two routes  were to be $T_3^A\neq T_3^B$,
it would imply by conservation of energy that either the system or the `reservoir' have negative heat capacity which is of course impossible. 
Because the system and the `reservoir' constitute our `universe', their final equilibrium temperature and their final states are
uniquely defined.

Therefore, since both energy and entropy are functions of state, we necessarily have that 
\bmath
\beq
 Q_A=Q_B
 \eeq
 and
 \beq \Delta S_{univ, A}=\Delta S_{univ, B}
 \eeq
 \emath
From Eq. (5a) , we learn that the Joule heat is given by 
\bmath
\beq
Q_J=L(T_1)-L(T_2)+ \int_{T_2}^{T_1} dT [C_s(T)-C_n(T)]
\eeq
and  from Eq. (5b) we obtain,  using Eq. (6a),  that
\beq
\frac{L(T_2)}{T_2}-\frac{L(T_1)}{T_1}= \int_{T_2}^{T_1} dT \frac{ [C_s(T)-C_n(T)]}{T} .
\eeq
\emath
Since the latent heat is given by $L(T)=T(S_n(T)-S_s(T))$, where $S_n$ and $S_s$ are entropies in the normal and
superconducting states, and $C_{s,n}(T)=T(\partial S_{s,n}/\partial T)$, Eq. (6b) is true. This demonstrates the
consistency of our approach.  

From Eq. (6a) we obtain, using these definitions and integrating by parts 
that the Joule heat is simply given by
\beq
Q_J= \int_{T_2}^{T_1} dT [S_n(T)-S_s(T)] .
\eeq
in a process where the system in the normal state is supercooled from the equilibrium transition temperature $T_1$ to
a lower temperature $T_2$. 
One way to understand this result is that when the system is supercooled it accumulates extra entropy by staying in  the normal state
relative to what it would have in the superconducting state, and rids itself
of this extra entropy when it undergoes the transition at the lower temperature by generation of Joule heat.

             \begin{figure}
 \resizebox{8.5cm}{!}{\includegraphics[width=6cm]{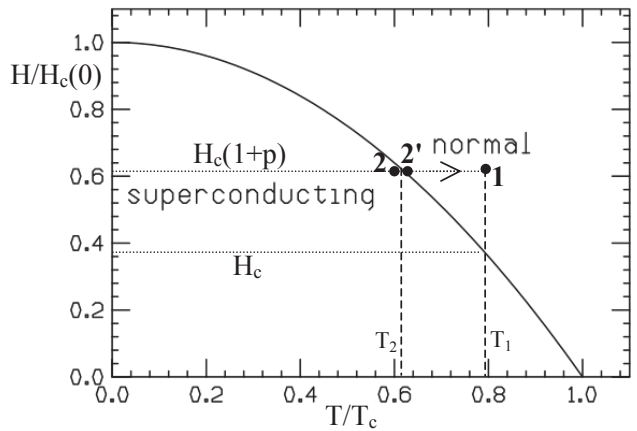}} 
 \caption { S-N transition: states 2 and  2' are  superconducting and normal states on the coexistence curve at temperature $T_2$, state 1 is the normal state at higher
 temperature $T_1$, all in external field $H_c(T_2) \equiv H_c(1+p)$.  The critical field at temperature $T_1$ is $H_c \equiv H_c(T_1)$.
  States 2, 2', and 1 are equilibrium states.
  The superheated non-equilibrium state 1' is the  superconducting state for the same temperature and magnetic field as the equilibrium normal state 1.
    The system interchanges heat with a heat reservoir at temperature $T_1$.
  }
 \label{figure1}
 \end{figure} 
 
In a similar fashion we can analyze the processes in the superconductor to normal  transition shown in Fig. 5: here, route A involves the equilibrium
transition between superconducting state 2 and normal state 2' at coexistence, and route B involves superheating the system in the 
superconducting state to a non-equilibrium state 1' at temperature $T_1$, then undergoing the transition to the normal
state 1  in a finite amount of time absorbing latent heat from a reservoir at temperature $T_1$ and generating Joule heat. 
The corresponding equations are
(here, $Q_{A,B}$ are the heats absorbed by the system):
\beq
Q_A=L(T_2)+\int_{T_2}^{T_1} dT C_n(T)  .
\eeq
\beq
\Delta S_{univ, A}=L(T_2)(\frac{1}{T_2}-\frac{1}{T_1})+\int_{T_2}^{T_1} dT C_n(T)[\frac{1}{T}-\frac{1}{T_1}]  
\eeq
\beq
Q_B= \int_{T_2}^{T_1} dT C_s(T) +L(T_1)-Q_J
\eeq
\beq
\Delta S_{univ, B}=   \int_{T_2}^{T_1} dT C_s(T)[\frac{1}{T}-\frac{1}{T_1}]   +\frac{Q_J}{T_1}
\eeq
leading through the conditions Eq. (5)  to 
\bmath
\beq
Q_J=L(T_1)-L(T_2)+ \int_{T_2}^{T_1} dT [C_s(T)-C_n(T)]
\eeq
and
\beq
\frac{L(T_2)}{T_2}-\frac{L(T_1)}{T_1}= \int_{T_2}^{T_1} dT \frac{ [C_s(T)-C_n(T)]}{T} .
\eeq
\emath
which are identical to Eqs. (6). Therefore, again Eq. (12b) demonstrates the consistency of our approach also for superheating. Eq. (12a) implies
that also for superheating the Joule heat generated is given by the same Eq. (7)
\beq \tag{7}
Q_J= \int_{T_2}^{T_1} dT [S_n(T)-S_s(T)]  
\eeq
in a process where the system in the superconducting state is superheated to a temperature higher than the coexistence
temperature and then undergoes the transition to the normal state.
This is more difficult to understand intuitively than for the case of supercooling, but equally true. 
 
\section{Clausius-Clapeyron relation and Joule heat}

We can shed further light on these results by considering the equation for the coexistence curve $H_c(T)$. 
The Clausius-Clapeyron equation for a system with thermodynamic variables $T$, $V$, $P$ ($V$=volume, $P$=pressure) undergoing a first order phase transformation is well known:
   \beq
   \frac{dP}{dT}=\frac{L(T)}{T\Delta V}
   \eeq
   where $L$ is the latent heat and $\Delta V$ the volume change. The analogous equation for a superconductor is \cite{reif}
   \beq
   \frac{dH_c}{dT}=\frac{L(T)}{T(M_s-M_n)}
   \eeq
   where $M_n=0$ is the magnetization in the normal state and $M_s=-H_c/(4\pi)$ is the magnetization in the superconducting state.
   From Eq. (14)  it follows that
   \beq
   \frac{L(T)}{T}=-\frac{H_c}{4\pi}\frac{dH_c}{dT}
   \eeq
   hence
      \beq
S_n(T)-S_s(T)=-\frac{H_c}{4\pi}\frac{dH_c}{dT} .
   \eeq
   Replacing in Eq. (7) we find that 
 for both supercooling and superheating the Joule heat generated in the transition is given by
   \beq
   Q_J=\frac{H_c(T_2)^2}{8\pi}-\frac{H_c(T_1)^2}{8\pi}
   \eeq
   which is simply the difference in the condensation free energies at the lower and the higher temperature.

   \section{Electrodynamics}

To check  our result Eq. (17) we consider the electromagnetic energy equation
\beq
\frac{d}{dt}(\frac{H^2}{8\pi}) = -\vec{J}\cdot \vec{E} - \frac{c}{4\pi} \vec{\nabla}\cdot (\vec{E}\times \vec{H}),
\eeq
 first for the N-S transition, where the system makes the transition from normal to superconducting in an applied
magnetic field $H_c(1-p)=H_c(T_1)$ at the supercooled temperature $T_2$. 
 The left side represents the change in energy of the electromagnetic field as the magnetic field  
 is expelled from  the body, the first term on the right side is the work done by the electromagnetic
 field on  currents in this process, and the second term is the outflow of electromagnetic energy. Integrating over the volume of the body $V$  and over time we find for the
 change in electromagnetic energy per unit volume
 \beq
\frac{1}{V} \int d^3r \int_0^\infty dt \frac{d}{dt}(\frac{H^2}{8\pi}) =-\frac{H_c(T_1)^2 }{8 \pi} .
 \eeq
 since at the end the initial magnetic field $H_c(T_1)$  is completely excluded from the body.
 From Faraday's law and assuming cylindrical symmetry we have for the electric field generated by the changing magnetic flux at the surface
 of the cylinder
 \beq
 \vec{E}(a,t)=-\frac{1}{2\pi a c}\frac{d}{dt}  \phi(t) \hat{\theta}
 \eeq
 where $a$ is the radius of the cylinder and $\phi(t)$ is the magnetic flux throught the cylinder, with
  $\phi(t=0)=\pi a ^2 H_c(T_1)$, $\phi(t=\infty)=0$. 
Integration of the second term on the right in Eq. (18), the energy outflow, over space and  time,
converting the volume integral to an integral over the surface   of the cylinder, 
using that $H=H_c(1-p)$ at the surface of the cylinder independent of time and Eq. (20)
for the electric field at the surface yields
 \beq
 \frac{1}{V} \int_0^\infty dt    \oint (- \frac{c}{4\pi}) (\vec{E}\times \vec{H}) \cdot d\vec{S}=-\frac{H_c(T_1)^2}{4 \pi} .
 \eeq
This gives the total electromagnetic energy flowing out through the surface of the sample during the
 transition.
 
 The current $\vec{J}$  in Eq. (18) flows in the azimuthal direction and is given by the sum of superconducting and normal currents
 \beq
J(r)=J_s(r)+J_n(r)
\eeq
where $J_s(r)$ flows in the region $r\leq r_0(t)$ and is of appreciable magnitude only within $\lambda_L$ of the phase boundary, where $\lambda_L$
is the London penetration depth. $r_0(t)$ is the radius of the phase boundary  at time $t$. 
It is important to note  the fact that at the superconductor-normal phase boundary the magnetic field
is given by $H_c(T_2)\equiv H_c$, as indicated in Fig. 1, since there is coexistence of superconducting and normal phases
at that radius \cite{pippard}. 
The extra magnetic field relative to the field at the cylinder surface is supplied by the current $J_n$  induced by the Faraday field  flowing in the normal region $r\geq r_0(t)$ \cite{pippard}.
Integration of the second term in Eq. (18) over the superconducting current yields \cite{pippardmine}
\beq
\frac{1}{V} \int d^3r \int_0^\infty dt  (-\vec{J}_s \cdot \vec{E} )=  \frac{H_c(T_2)^2}{8\pi}  .
\eeq
This is because the Faraday field {\it decelerates} the supercurrent \cite{pippardmine} as the phase boundary
moves out .

The Joule heat per unit volume generated during the transition is
\beq
Q_J \equiv \frac{1}{V} \int d^3r \int_0^\infty dt  \vec{J}_n \cdot \vec{E}
\eeq
hence from integrating Eq. (18) over space and time using Eqs.  (19), (21), (22) and (23) we have
\beq
-\frac{H_c(T_1)^2}{8\pi}= \frac{H_c(T_2)^2}{8\pi} - Q_J - \frac{H_c(T_1)}{4\pi}
\eeq
which implies 
   \beq
   Q_J=\frac{H_c(T_2)^2}{8\pi}-\frac{H_c(T_1)^2}{8\pi}
   \eeq
   identical to the thermodynamic result Eq. (17).
    
   We leave it as a simple  exercise for the reader to derive the same equation for the case of superheating, where 
   the electromagnetic energy flows into the system through the surface and the Faraday electric field
   speeds up rather than slows down the supercurrent.

 \section{direct calculation of the  Joule heat}
 We next calculate the Joule heat generated in the process shown in Fig. 1 directly, assuming it originates $only$ from
 current in the normal region.  Ampere's law and Faraday's law in cylindrical geometry  yield
 \bmath
  \beq
 \frac{\partial H}{\partial r}=-\frac{4\pi}{c}J
 \eeq
  \beq
 \frac{1}{r} \frac{\partial}{\partial r}(rE_F)=-\frac{1}{c}\frac{\partial H}{\partial t}
 \eeq
 \emath
 with $\vec{H}=H\hat{z}$, $\vec{J}=J\hat{\theta}$ and $\vec{E}=E_F \hat{\theta}$ the magnetic field in the $\hat{z}$ direction,
 current density and electric field in the azimuthal direction respectively. 
 The boundary conditions are
 \bmath
 \beq
 H(r=a)=H_c(1-p)
 \eeq
 \beq
 H(r=r_0)=H_c .
 \eeq
 \emath
 
 As pointed out by Pippard in his seminal paper \cite{pippard}, these equations cannot be solved exactly but can be solved in a 
 power series expansion in $p$. To lowest order in $p$ we may assume that the magnetic field is $H_c$ for all $r\geq r_0$, hence
 the Faraday electric field is given for $r\geq r_0$  by
 \beq
 E_F(r)=\frac{r_0}{r}\frac{\dot{r}_0}{c}H_c
 \eeq
 In the normal region $r\geq r_0$ we assume the normal current $J_n$ obeys the constitutive relation
 \beq
 J_n(r)=\sigma_n(r) E_F(r) .
 \eeq
We allow the normal conductivity $\sigma_n$ to depend on $r$ for generality. From Eqs. (30), (29) and (27a)
we deduce
\beq
\frac{\partial H}{\partial r}=-\frac{4\pi}{c^2} \sigma_n(r)\frac{r_0}{r} \dot{r}_0
\eeq
and integrating between $r=r_0$ and $r=a$ and using Eq. (28) we obtain
\beq
pH_c=\frac{4\pi}{c^2} r_0 \dot{r}_0 H_c \int_{r_0}^a dr \frac{\sigma_n(r)}{r} .
\eeq
The normal current generates Joule heat per unit volume at rate given by
\beq
\frac{\partial w}{\partial t}=J_n(r)  E_F(r)=\sigma_n(r) E_F(r)^2 .
\eeq
Integrating over the volume of the normal metal the rate of Joule heat generation  is, using Eq. (29)
\beq
\frac{\partial W}{\partial t}=\int d^3 r \frac{\partial w}{\partial t}=
2\pi \int _{r_0}^a dr r \sigma_n(r) \frac{r_0^2}{r^2}\frac{\dot{r}_0^2}{c^2}H_c^2
\eeq
and using Eq. (32) we obtain the simple result
\beq
\frac{\partial W}{\partial t}=\frac{1}{2}pH_c^2 r_0\dot{r}_0 .
\eeq
Finally, integrating Eq. (35) over time and dividing by the volume of the cylinder we obtain the  Joule heat per unit volume generated {\it in the 
normal region} during the entire
process:
\beq
Q_J=\frac{1}{\pi a^2} \int_0^\infty dt \frac{\partial W}{\partial t} = \frac{H_c^2}{4\pi} p .
\eeq
Now from Eq. (26) we have, with $H_c(T_2)=H_c$, $H_c(T_1)=H_c(1-p)$
\beq
Q_J=\frac{H_c^2}{8\pi}-\frac{(H_c(1-p))^2}{8\pi} = \frac{H_c^2}{4\pi} p + O(p^2)
\eeq
in agreement with Eq. (36) to lowest order in p.

We can in fact obtain a more accurate answer by taking into account the fact that the magnetic field
changes between $H_c$ and $H_c(1-p)$ in the region $r_0\leq r \leq a$. In Eq. (33), we replace the normal
current in terms of the magnetic field using Eq. (27a)
 \beq
\frac{\partial w}{\partial t}=-\frac{c}{4\pi} \frac{\partial H}{\partial r}  E_F(r)
\eeq
and for the Faraday field we use Eq. (29) replacing $H_c$ by the average field in the region $r_0\leq r \leq a$, 
$H_c(1-p/2)$
 \beq
\frac{\partial w}{\partial t}=-\frac{c}{4\pi} \frac{\partial H}{\partial r}  \frac{r_0}{r}\frac{\dot{r}_0}{c}H_c(1-\frac{p}{2})
\eeq
from which we obtain integrating over space and time
\beq
Q_J=\frac{H_c^2}{4\pi}p(1-\frac{p}{2})
\eeq
in exact agreement with Eq. (37).

The equality of Eqs. (36) or (40), calculated using only the current in the normal region, with   Eqs. (17) and (26)
obtained from thermodynamics and electrodynamics,   implies that {\it no} Joule heat was generated in the superconducting region in this process.
The same result is obtained by considering the Joule heat generated in the normal region  in the S-N  transition
in the presence of magnetic field $H_c(1+p)\equiv H_c(T_2)$ (Fig. 5).
We discuss the significance of these results in what follows.
 
\section{The missing Joule heat}
We consider for definiteness the process of supercooling, the same issues arise for superheating.
The magnetic field does not drop to zero discontinuously  at the phase boundary $r=r_0$, rather it decays smoothly as governed by
the London penetration depth. For $r\leq r_0$ London's equation applies:
\beq
\vec{J_s}(r,t)=-\frac{c}{4\pi \lambda_L^2}\vec{A}(r,t)
\eeq
with $\vec{A}$ the magnetic vector potential, given by  (assuming $r_0>>\lambda_L$)  \cite{pippardmine}
\beq
\vec{A}(r,t)=H_c \lambda_Le^{(r-r_0)/\lambda_L} \hat{\theta} .
\eeq
The magnetic field $\vec{H}=\vec{\nabla}\times \vec{A}$ is 
\beq
\vec{H}(r)=H_c e^{(r-r_0)/\lambda_L}\hat{z}
\eeq
and the  Faraday electric field $\vec{E}(r,t)=-(1/c)\partial \vec{A}(r,t)/\partial t$ is
\beq
E(r)=\frac{\dot{r}_0}{c}H_c e^{(r-r_0)/\lambda_L} .
\eeq

Within the conventional theory of superconductivity, at finite temperatures a superconductor can be modeled
approximately as a two-fluid model \cite{tinkham2f}, with normal and superconducting electrons of density
$n_n$, $n_s=n-n_n$, with $n$ the conduction electron density. $n_s$ and $n_n$ depend on temperature.
A more detailed  treatment using Bogoliubov quasiparticles as the normal state excitations would yield equivalent results. 
The electric field Eq. (44) will give rise to a {\it normal current}
\beq
J'_n(r)=\sigma_0 E(r) .
\eeq
where we have approximately $\sigma _0 =(n_n/n)\sigma_n$. The predicted  rate of Joule heat generation in the superconducting  region is then
\beq
\frac{\partial w_s }{\partial t}=\sigma_0 E(r)^2=\sigma_0 \frac{\dot{r}_0^2}{c^2}H_c ^2 e^{2(r-r_0)/\lambda_L} 
\eeq
and performing the spatial integral
\beq
\frac{\partial W_s}{\partial t}=\int_{r\leq r_0} d^3 r \frac{\partial w}{\partial t}=\pi \sigma_0 \frac{\dot{r}_0^2}{c^2}r_0 \lambda_L H_c^2 .
\eeq
Under the assumption that $\sigma_n$ is
independent of $r$ we can integrate Eq. (31) over space and time to obtain 
\beq
(\frac{r_0}{a})^2[1+2ln\frac{a}{r_0}]=\frac{t}{t_0} .
\eeq
where 
\beq
t_0=\frac{\pi \sigma_n a^2}{p c^2} 
\eeq
is the total time to expel the magnetic field. 
For simplicity we can assume $\dot{r}_0\sim a/t_0$ in Eq. (47), and performing the time integral we find
for the Joule heat per unit volume generated in the superconducting region
predicted by the conventional theory:
\beq
q\equiv \int d^3 r \frac{\partial W_s}{\partial t}= \frac{H_c^2}{\pi} p \frac{\sigma_0}{\sigma_n}\frac{\lambda_L}{a}
\eeq
or
\beq
q=4\frac{\sigma_0}{\sigma_n}\frac{\lambda_L}{a} Q_J .
\eeq

As one would expect, this Joule heat $q$ is proportional to the London penetration depth $\lambda_L$.
But we saw in Sects. II-IV that the total Joule heat depends only on thermodynamic properties and
is independent of $\lambda_L$. And we saw in Sect. V that the Joule heat generated in the normal region
accounts for the entire Joule heat predicted by thermodynamics and electrodynamics. Therefore we conclude that 
$q$ does not exist. Therefore there cannot be a
normal azimuthal current induced by the Faraday field in the superconducting region, contrary to what Eq. (45) says.

\section{other considerations}

There are in fact other reasons for  why a normal current in the superconducting region cannot exist. 
The total current in the superconducting region is fixed by the fact that it has to nullify the magnetic field in the deep interior.
The supercurrent $J_s$ in the superconducting region  is given by
\beq
\vec{J}_s=-\frac{c}{4\pi \lambda_L}H_c e^{(r-r_0)/\lambda_L} \hat{\theta}
\eeq
as follows from the London equation Eq. (41), 
in the absence of normal current. If a normal current $J'_n$ (Eq. (45)) were to be induced in the boundary layer by the Faraday field, it would require
that an additional supercurrent 
\beq
\vec{J}'_s=-\vec{J}'_n
\eeq
 be generated in order not to change the total current, as indicated schematically in Figs. 1 and 2, so as to  keep the interior magnetic field equal to zero. 
 The total supercurrent $\vec{J}_s+\vec{J}_s'$ would no longer satisfy London's equation Eq. (41). This would be  in disagreement with the conventional theory of
 superconductivity where the London equation  is a consequence of the fact that the canonical momentum of electrons in the supercurrent is zero in a simply connected
 geometry.
 
Furthermore, consider the energy of the currents. 
The kinetic energy of the supercurrent $J_s$ per unit volume   is given by
\beq
K_s(r)=\frac{m_e}{2n_s e^2}J_s(r) ^2
\eeq
where $m_e$ is the electron mass (we ignore possible differences between bare and effective mass for simplicity \cite{mstar}). The London
penetration depth $\lambda_L$ satisfies \cite{tinkham}
\beq
\frac{1}{\lambda_L^2}=\frac{4\pi n_s e^2}{m_e c^2}
\eeq
and Eqs. (52), (54)  and  (55) yield for the kinetic energy of the supercurrent at the phase boundary
\beq 
K_s(r_0)=\frac{H_c^2}{8\pi}
\eeq
which is the condition for phase equilibrium at the normal-superconductor boundary, first discussed by H. London \cite{londonh}.
As electrons condense into the superconducting state, their condensation energy $H_c^2/8\pi$ provides precisely the kinetic energy necessary for the
electrons to join the supercurrent at the phase boundary.

Instead, if there is the additional supercurrent Eq. (53), the total kinetic energy of the supercurrent at the phase boundary would be different.
Consider first the N-S transition, Fig. 1. The extra supercurrent $\vec{J}'_s$ flows in the same direction as $\vec{J}_s$, hence the
kinetic energy of the supercurrent would be 
\beq
K_{s}(r_0)=\frac{m_e}{2n_s}(J_s(r_0)+J'_{s}(r_0))^2>\frac{H_c^2}{8\pi} .
\eeq
In other words, electrons condensing into the superconducting state would have to acquire a kinetic energy larger than the condensation energy. That is impossible.
There is no extra energy source to supply the kinetic energy associated with the extra supercurrent nor with the normal current $J_n'$.
For the S-N transition, the extra supercurrent $\vec{J}'_s$ flows in direction opposite to $\vec{J}_s$, here the kinetic energy of the
supercurrent would be 
\beq
K_{s}(r_0)=\frac{m_e}{2n_s}(J_s(r_0)-J'_{s}(r_0))^2<\frac{H_c^2}{8\pi} .
\eeq
Still, the sum of it plus the kinetic energy associated with $J_n'$ would not equal the available energy $ H_c^2/(8\pi)$. 
Also the lack of symmetry between the supercooled and superheated situations, contrary to the symmetry found in the previous
sections,  indicates that the term
$J_s'$ should not be in either Eq. (57) or (58), hence that $J_n'=0$.

 \section{relation with earlier work}
 Note that in our discussion of route A in the cooling process, Figs. 3 and 4, we computed the entropy change 
 $\Delta S_{univ,A}$, Eq. (2), under the assumption that no Joule heat is generated when the system is cooled in the superconducting
 state from temperature $T_1$ to $T_2$, states 1' to 2. We did not make any assumption about the rate at which this cooling
 occurs. During this process, the London penetration depth decreases and a Faraday electric field is induced
 within $\lambda_L$ of the surface of the cylinder. We analyzed that process in ref. \cite{inconsistency} and pointed out
 that no normal current can be induced by the Faraday field during that process because that would be incompatible with
 thermodynamics. That is consistent with what we assumed here in Eq. (2), and with what we concluded in 
 Sects. VI and VII regarding normal current in the superconducting region within $\lambda_L$ of the phase boundary
 where the Faraday field exists: there isn't any.
 
 Furthermore, within the conventional theory, additional normal current is generated in the normal region near the phase boundary
 to compensate for the momentum acquired or lost by electrons entering or leaving the superconducting state.
 We analyzed this in Ref. \cite{entropy}, where we found that this additional normal current generates
 Joule heat given by (per unit volume)
 \beq
 Q_n=\frac{1}{\pi a^2}\int_0^{t_0} \frac{dt}{\tau} (2\pi r_0 \ell) \frac{H_c^2}{8\pi}(\frac{\dot{r}_0}{v_F})^2
 \eeq
 with $\ell$ the mean free path, $\tau$ the collision time and $v_F=\ell/\tau$ the Fermi velocity. The speed of motion
 of the phase boundary is given by
 \beq
 \dot{r}_0=\frac{pc^2}{4\pi \sigma_n r_0 ln(a/r_0)}
 \eeq
 Assuming for simplicity that $\dot{r}_0$ is approximately  time independent, Eq. (59) yields simply
 \beq
 Q_n= \frac{H_c^2}{8\pi}\frac{\dot{r}_0}{v_F} .
 \eeq
 In fact, more detailed arguments   \cite{entropy} taking into account the distribution in phase space of the dissociating or condensing electrons show that $Q_n$ is larger than Eq. (61) by a large numerical factor
 $((8/3) (k_F \lambda_L))$, with $k_F$ the Fermi momentum.
 
 In the present context, the important point about the Joule heat Eq. (61) is that it is proportional 
 to the speed of the process,
 $\dot{r}_0$.   $\dot{r}_0$
  is approximately given by
 \beq
 \dot{r}_0\sim \frac{a}{t_0}=\frac{pc^2}{\pi\sigma_n a}
 \eeq
 so in particular it depends on the normal state conductivity $\sigma_n$ and on the radius of the sample $a$. 
Additionally,  $Q_n$  depends  also  on $v_F$, the Fermi velocity. 
 However, according to our results in the earlier sections the total Joule heat $Q_J$ is
 independent of $all$  those variables. This confirms the conclusion already reached in Ref. \cite{entropy} that
 the prediction of the conventional theory that normal current is generated in the process where electrons 
 go from the normal to the superconducting state or from the superconducting to the normal state to 
 satisfy momentum conservation \cite{halperin}, cannot be correct.

\section{summary and discussion}
We have   calculated  the Joule heat per unit volume  that is generated when a normal
metal expels a magnetic field in the transition to the superconducting state, and when a superconductor goes normal in the 
presence of a magnetic field. We have found from purely thermodynamic considerations  that the Joule heat takes the same simple form in both cases, Eq. (7) or
equivalently Eq. (17), 
independent of the normal state conductivity and of  the time that the process takes. 
This result was corroborated by a calculation using purely electrodynamic considerations, Eq. (26). To our knowledge, the fact 
that the Joule heat in these transitions is simply related to  thermodynamic properties has not been pointed out before.

At first sight the  result may seem counterintuitive. It follows from the fact that the time the process takes is proportional to the normal
state conductivity $\sigma_n$, as given by Eq. (49). For large $\sigma_n$ the process occurs very slowly and the Faraday
field is very small, for small $\sigma_n$ the process is fast and the Faraday field is large, but the total Joule heat generated is
the same in all cases.
The physical reason that the process is slow if the normal state conductivity is large is the following: what limits the speed of the
process is that the magnetic field at the phase boundary is exactly $H_c$. The role of the current in the normal region in the cooling process  is to 
generate the extra magnetic field $(pH_c)$ to increase the magnetic field from its value at the cylinder surface to its value at the
phase boundary, or in the heating process to reduce the applied magnetic field from its value at the cylinder surface by $(pH_c)$. If the process proceeds too fast, the induced normal current would produce a magnetic field at the phase 
boundary larger than $H_c$ in the cooling process or smaller than $H_c$ in the heating process, reversing the direction of the process.

Note however that the rate at which we cool from state 1' to state 2 in route A is $not$ determined by
$\sigma_n$, but rather by the conditions of the experiment, as discussed in ref. \cite{inconsistency}.
The conventional theory would predict that the Joule heat generated in that process varies with the
cooling rate, while in fact $no$ Joule heat can be generated in this process independent of the
cooling rate to be consistent
with thermodynamics \cite{inconsistency}.

We then showed by direct calculation that the Joule heat predicted by thermodynamics results from the normal current induced by the Faraday
electric field {\it in the normal region only}.
However, an induced Faraday  electric field necessarily exists close to the phase boundary in the superconducting region during these processes.
Within the conventional theory of superconductivity  this electric field will both affect 
superfluid and normal electrons (i.e. Bogoliubov quasiparticles). For the superfluid electrons, in the cooling process the Faraday field  slows them down so that as the phase
boundary moves further out  and they become part of the interior their velocity slows to zero, in the heating process it speeds them up
so that as the phase boundary moves in the velocity  reaches the value necessary to generate the current $J_s$ at the boundary, as shown in 
\cite{pippardmine}. For the normal electrons, the Faraday field  will generate a normal current and associated Joule heat Eq. (51) within the
conventional theory, in an amount proportional to the London penetration depth $\lambda_L$. But this is impossible, since all the Joule heat allowed by thermodynamics as well as by electrodynamics is  generated in the normal region, and in
addition is independent of $\lambda_L$.

Therefore, within the conventional theory, two superconductors with the same thermodynamic properties but different transport properties would
dissipate different amounts of Joule heat in the same thermodynamic process. For example, if one of them had more impurities it could have
a much larger London penetration depth \cite{tinkham} with little or no change in its thermodynamic properties, hence a much larger boundary layer where
Joule heat would be dissipated according to the conventional theory, in contradiction with thermodynamics.

In addition, we pointed out in Sect. VIII  that within the conventional theory the need to conserve momentum
in the process of conversion between superfluid and normal
electrons generates additional Joule heat and entropy \cite{entropy} per unit volume, in an amount that
depends on $\dot{r}_0, a, v_F, k_F, \lambda_L, \sigma_n$:  the speed of the process, the radius of the sample, the Fermi velocity, the Fermi momentum, the London penetration depth, and the
normal state conductivity, six different variables, of which at most one can be expressed in terms of the others. 
Yet the total Joule heat Eq. (7) does not  depend on $any$ of these variables.

We conclude from these considerations that there is a problem with the conventional theory of superconductivity.  The Faraday electric field
does not induce a normal current in the superconducting region in these processes, contrary to what the conventional theory of superconductivity
predicts \cite{tinkham}. We reached the same conclusion in recent work where we
considered processes where the temperature changes always below $T_c$: in such processes a Faraday electric field exists because the 
London penetration depth and consequently the magnetic flux is changing.  What these processes have in common
with the ones discussed in this paper is that the density of superfluid electrons is changing. We conclude that in such situations the resulting electric field does not generate a normal current,
contrary to situations where an electric field is produced by ac currents or electromagnetic waves, where normal current is known to be
generated \cite{tinkham2f}.   Neither is a normal current generated
in the process of normal-superconductor conversion near the interface, as discussed in Ref. \cite{entropy},
contrary to what the conventional theory predicts.

In contrast to the conventional theory,   within the theory of hole superconductivity \cite{holesc}  there is {\it radial motion} of charge in the normal-superconductor transition
in a magnetic field. We have argued that this is necessary to explain the process of magnetic field expulsion and how
momentum is transferred from electrons to the body as a whole in a reversible way to account for momentum conservation \cite{momentum, entropy}.
This physics can also explain the absence of
Joule heat  in the boundary layer in the situation discussed in this paper, as well as the
absence of Joule heat   in a process where the temperature below $T_c$ is
changed \cite{inconsistency}. Note that a radial normal current in the presence of an azimuthal electric field does not give rise to dissipation. This physics   also explains the absence of Joule heat and entropy generation forbidden by thermodynamics when momentum
is transferred between electrons and the body as a whole as discussed in \cite{entropy}. It requires the 
charge carriers in the normal state to be holes \cite{whyholes}.

  \acknowledgements
 The author is grateful to Bert Halperin and  Tony Leggett for helpful discussions.

 \end{document}